\documentclass[aps,twocolumn,prl,floatfix,superscriptaddress,showpacs,longbilbiography]{revtex4-1}
\usepackage[utf8]{inputenc} 
\usepackage{amssymb}

\usepackage{graphicx}
\usepackage{dcolumn}
\usepackage{braket}
\usepackage{bm}
\usepackage{amsfonts}
\usepackage{amsmath}
\usepackage{amssymb}
\usepackage{color,soul}
\usepackage{wasysym}
\usepackage{mathrsfs}
\usepackage{times}
\usepackage[dvipsnames,svgnames,table]{xcolor}
\usepackage{ulem}
\usepackage{hyperref}
\usepackage{fancyhdr}
\usepackage{float}
\usepackage[english]{babel}
\usepackage{hyperref}
\usepackage[caption=false]{subfig}

\hypersetup{
pdfstartview={FitH},
pdftitle={2017},    
pdfauthor={LEFFT},  
colorlinks=true,    
linkcolor=NavyBlue, 
citecolor=Maroon,   
filecolor=NavyBlue, 
urlcolor=NavyBlue   
}

\newcommand{\bea}{\begin{eqnarray}}
\newcommand{\eea}{\end{eqnarray}}

\begin{document}
\title{Non-Hermitian robust edge states in one-dimension:\\Anomalous localization and eigenspace condensation at exceptional points}

\author{V. M. Martinez Alvarez}
\affiliation{Laborat\'orio de F\'{\i}sica Te\'orica e Computacional, Departamento de F\'{\i}sica, Universidade Federal de Pernambuco, Recife 50670-901, PE, Brazil}
\author{J. E. Barrios Vargas}
\affiliation{Departamento de F\'{\i}sica, Facultad de Ciencias F\'{\i}sicas y Matem\'aticas, Universidad de Chile, Santiago, Chile}
\author{L. E. F. Foa Torres}
\affiliation{Departamento de F\'{\i}sica, Facultad de Ciencias F\'{\i}sicas y Matem\'aticas, Universidad de Chile, Santiago, Chile}

\begin{abstract}

Capital to topological insulators, the bulk-boundary correspondence ties a topological invariant computed from the bulk (extended) states with those at the boundary, which are hence robust to disorder. Here we put forward an ordering unique to non-Hermitian lattices, whereby a pristine system becomes devoid of extended states, a property which turns out to be robust to disorder. This is enabled by a peculiar type of non-Hermitian degeneracy where a macroscopic fraction of the states coalesce at a single point with geometrical multiplicity of $1$, that we call a \textit{phenomenal point}. 

\end{abstract}

\date{\today}
\maketitle

\textit{Introduction.--} More than 30 years ago, the pioneering studies on the topological origin of the quantum Hall effect with ~\cite{thouless_quantized_1982} and without Landau levels~\cite{haldane_model_1988}, sparked a revolution that reshaped condensed matter physics. This reached a decisive moment when new insights~\cite{kane_quantum_2005,bernevig_quantum_2006} ultimately led to the discovery of topological insulators in two~\cite{konig_quantum_2007} and three dimensions~\cite{hsieh_topological_2008}. Now, besides the  accelerating advance on these already exciting materials, a plethora of new trends is building momentum: From the search of gapless but topological phases such as the Weyl semimetals~\cite{vafek_dirac_2014}, to new topological states in systems out of equilibrium ~\cite{oka_photovoltaic_2009,lindner_floquet_2011,perez-piskunow_floquet_2014} and non-Hermitian lattices~\cite{rudner_topological_2009,diehl_topology_2011,yuce_topological_2015,san-jose_majorana_2016,lee_anomalous_2016,leykam_edge_2017}. 

Whether or not non-Hermitian systems~\cite{rudner_topological_2009,diehl_topology_2011,yuce_topological_2015,san-jose_majorana_2016,lee_anomalous_2016,leykam_edge_2017} can bear nontrivial and stable 
topological states and, ultimately, the existence of a main guiding principle, i.e. a bulk-boundary correspondence, are issues at the center of an intense debate and controversy~\cite{hu_absence_2011,esaki_edge_2011,lee_anomalous_2016,xiong_why_2017,Shen_topological_2017}. One of such fascinating focus of discussion has been inspired by a one-dimensional (1D) model where a non-Hermitian Hamiltonian is built as to encircle an exceptional point (EP) in momentum space~\cite{lee_anomalous_2016}. Exceptional points are singularities in non-Hermitian systems, where not only the eigenvalues but also the eigenfunctions coalesce, thus making the Hamiltonian non diagonalizable (defective)~\cite{heiss_physics_2012}. EPs, which can also be of higher-order~\cite{Graefe2008,Eleuch2016,Jing2017,Pick17,Hodaei2017} as in this work, lead to intriguing phenomena such as unidirectional invisibility~\cite{Lin2011}, single-mode lasers ~\cite{Feng2014,Hodaei2014}, or enhanced sensitivity in optics~\cite{Chen2017,Hodaei2017,Rechtsman2017}. Curiously, although bearing a gapless bulk, the lattice model of Ref. ~\onlinecite{lee_anomalous_2016} has been shown to exhibit a \textit{single} zero-energy edge state localized on one side of the system, a fact that has been related to a fractional winding number of $1/2$. In contrast, other authors argue that the bulk-boundary correspondence altogether has to be abandoned in these non-Hermitian systems~\cite{xiong_why_2017}. Other recent studies add further interest to this exciting area~\cite{leykam_edge_2017,Shen_topological_2017,Kozii2017}.

In this Letter, we examine the emergence of an order unique to non-Hermitian systems. Our main conclusions are illustrated using the one-dimensional non-Hermitian lattice model of Ref.~\onlinecite{lee_anomalous_2016}. Beyond exceptional points, a macroscopic fraction of the states of a large system may coalesce at a single energy with geometrical multiplicity of $1$ (i.e. the eigenspace is spanned by a \textit{single} nonzero eigenvector) 
that we call a \textit{phenomenal point}. Interestingly, we show that, besides being dynamically stable (vanishing imaginary part of the eigenenergy), phenomenal points can have~\footnote{One must notice that not all phenomenal points lead to the localization properties mentioned in the text. This is the case for the ones in Ref.~\onlinecite{Graefe2008}, as they emerge in a system which in the infinite limit size cannot be assimilated to a translational invariant lattice as in this work.} an ordering effect including: a fully localized spectrum which remains fairly robust to disorder; and the orthogonality of left and right eigenstates (also called biorthogonality) which, in contrast with the case of exceptional points, can span a large range of system parameters~\footnote{We note that the large degeneracy and biorthogonality around phenomenal points breaks the assumptions underlying the classification presented in Ref.~\onlinecite{Shen_topological_2017}.}. Unlike usual Hermitian lattices where localization is typically tied to defects or disorder, here all states may be localized at the edge of a pristine lattice, thereby motivating the use of the word anomalous in our title.

We interpret the ordering effect near phenomenal points in terms of an environment-mediated interaction encoded in the non-Hermitian part of the Hamiltonian, that leads to the condensation of the eigenstates onto a few selected states. Although environment-mediated interaction effects have been studied theoretically and experimentally in the past~\cite{Bird2004,Alvarez2006,Yoon2012,Rotter2015}, in none of the mentioned cases the effect was as strong as in ours. Indeed, close to an EP two measures of non-Hermiticity, the Hamiltonian's defectiveness and the biorthogonality of its left and right eigenstates~\cite{moiseyev_non-hermitian_2011}, are typically very fragile, \textit{e.g.}, the branch point in the spectrum is removed by any small perturbation in the parameter space. In contrast, we find that around phenomenal points these strong fingerprints of non-Hermiticity may pervade over a wide range of parameter values with dramatic consequences.

\begin{figure}
\centering
\subfloat{\label{Fig_1a}\includegraphics[width=1.0\linewidth]{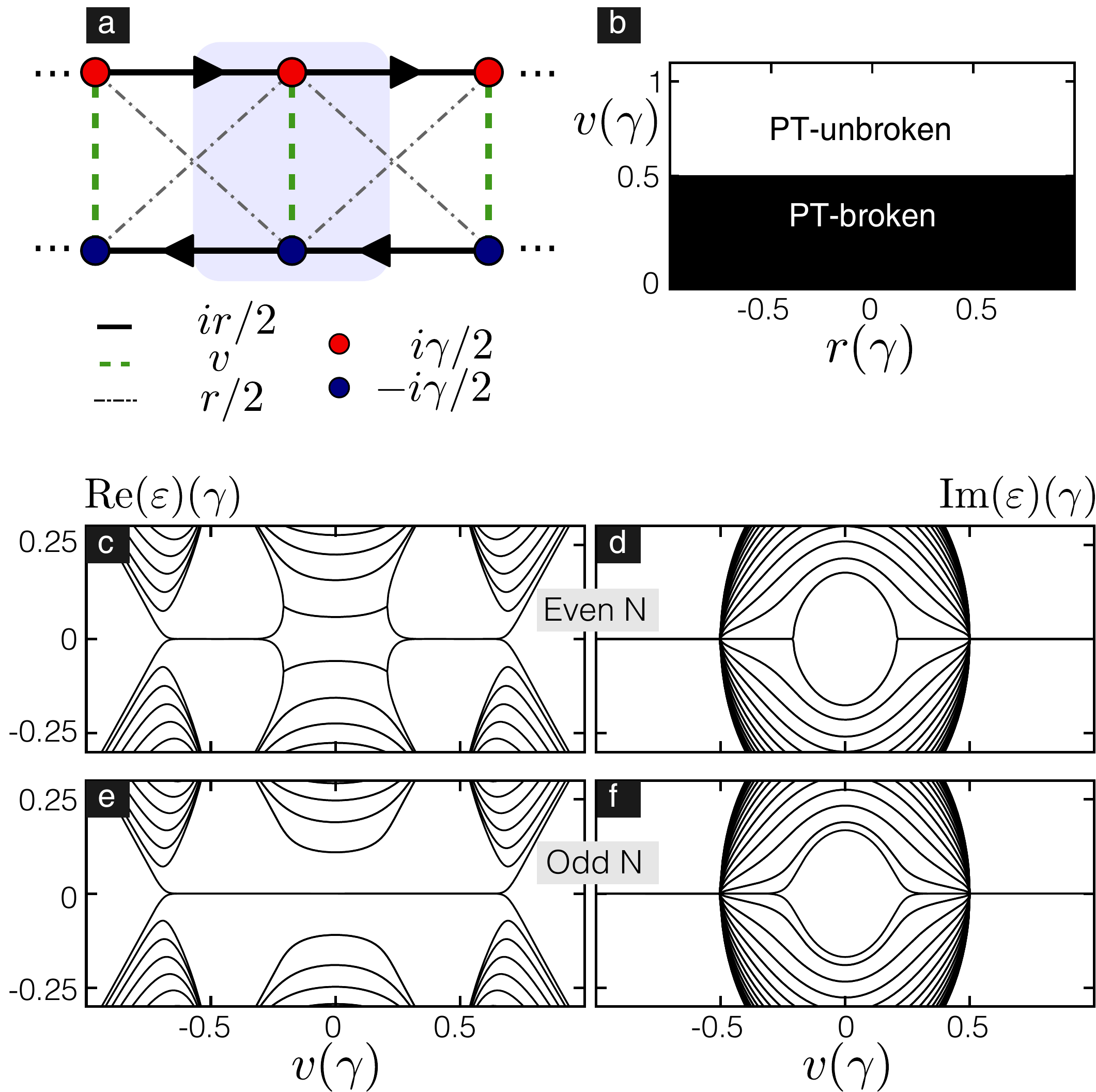}}
\subfloat{\label{Fig_1b}}
\subfloat{\label{Fig_1c}}
\subfloat{\label{Fig_1d}}
\subfloat{\label{Fig_1e}}
\subfloat{\label{Fig_1f}}
\caption{(Color online) (a) Scheme representing three cells of the lattice model introduced in the text. The imaginary on-site energies lead to the non-Hermiticity of the Hamiltonian. (b) Map of the PT-broken and unbroken regions as a function of the Hamiltonian parameters $v$ and $r$ (both in units of $\gamma$). (c) and (d) ((e) and (f)) show the real and imaginary parts of the spectrum obtained for a system of $N=30$ ($N=31$) unit cells and $r=0.5\gamma$.}
\label{Fig_1}
\end{figure}

\textit{Non-Hermitian model.--} To motivate our discussion let us re-examine the 1D non-Hermitian model of Ref.~\onlinecite{lee_anomalous_2016}:
\begin{equation}
{\cal H}_k=h_x(k)\sigma_x + \left(h_z(k) + \frac{i\gamma}{2}\right)\sigma_z,
\label{Hamiltonian}
\end{equation}
where $\sigma_x,\sigma_z$ are Pauli matrices, $\gamma$ is a real parameter tuning the degree of non-Hermiticity of ${\cal H}_k$ and $k$ is the wave-vector. $h_x(k)$ and $h_z(k)$ are chosen as to encircle the EP located at $(h_x,h_z)=(\pm \gamma/2,0)$. This is fulfilled by choosing: $h_x(k)=v+r\cos k$ and  $h_z(k)=r\sin k$, with $v$ and $r$ real parameters. Now, the model can be represented by a lattice with gains and losses on different sublattices, see Fig.~\ref{Fig_1}\subref{Fig_1a}. This model can be realized in an array of coupled
resonator optical waveguides~\cite{Hafezi2011} and in a photonic crystal~\cite{Weimann2017} as pointed out by Lee~\cite{lee_anomalous_2016}.

The Hamiltonian of Eq.~(\ref{Hamiltonian}) commutes with the composed parity-time, ${\cal PT}$, operator~\cite{lee_anomalous_2016}. PT-symmetry is said to be unbroken if all Hamiltonian eigenstates are also eigenstates of the ${\cal PT}$ operator, thereby having real-values eigenenergies. Fig.~\ref{Fig_1}\subref{Fig_1b} shows a phase diagram for the PT-broken and unbroken phases as a function of $v$ and $r$ for our model. We emphasize that although non-Hermitian Hamiltonians have extensively been discussed in the context of PT-invariant alternatives to Hermitian quantum  mechanics~\cite{Bender_prl_1998}, non-Hermitian Hamiltonians arise naturally either in the context of open quantum systems~\cite{rotter_non-hermitian_2009}, systems with gain and loss (as in photonics~\cite{ruter_observation_2010}) or because of the finite lifetime introduced by interactions~\cite{Kozii2017}.

\textit{Zero-energy edge states and parity effects.--} The bulk-boundary correspondence, which determines the existence of edge/surface states from bulk topological invariants, is a leitmotif ubiquitous in the field of topological insulators~\cite{hasan_colloquium_2010,ortmann_topological_2015,bernevig_topological_2013} and their more recent archetypes~\cite{rudner_anomalous_2013}. Much of the current debate revolves around the existence or not of such a relation in non-Hermitian systems and in this model in particular~\cite{lee_anomalous_2016,leykam_edge_2017,xiong_why_2017}. With this motivation we start our discussion by examining the zero-energy edge states in this model, though as we will show later on their relevance is lessened because of the presence of phenomenal points.

A finite section of the tight-binding lattice contains $N$ unit cells. $N$ may be even or odd. In Fig.~\ref{Fig_1}\subref{Fig_1c}-\subref{Fig_1f} we show the spectrum in the case of open boundary conditions for $r=0.5\gamma$ as a function of $v$ ($\gamma$ is taken as the unit of energy). When reaching $v=0.5\gamma$  from the right, two states coalesce at the EP into a single one with zero energy (and a vanishing imaginary part which makes it also dynamically stable). This state turns out to be localized at a single edge of the finite system and its existence has been attributed to a fractional winding number~\cite{lee_anomalous_2016}. Depending on the EP's chirality~\cite{heiss_physics_2012,heiss_chirality_2001}, the zero energy state turns out to be localized on one edge or the opposite. For $v<0.3\gamma$, the spectrum around $\Re(\varepsilon)\sim 0$ becomes distinct depending on the parity. Furthermore, for even $N$ the system encounters an additional EP as $v$ is lowered, the zero energy eigenvalue having algebraic (geometric) multiplicity $2$ ($1$) branches on two distinct states with non-vanishing energy (this does not happen for odd $N$ where the geometric multiplicity is kept until $v\sim 0$). 

An important question is how far away from the bulk's EP the existence of this state can be warranted. Fig.~\ref{Fig_1} gives a first hint, but let us now examine in more detail the localization and the robustness against disorder of the full spectrum.
\begin{figure}
\centering
\subfloat{\label{Fig_2a}\includegraphics[width=1.0\linewidth]{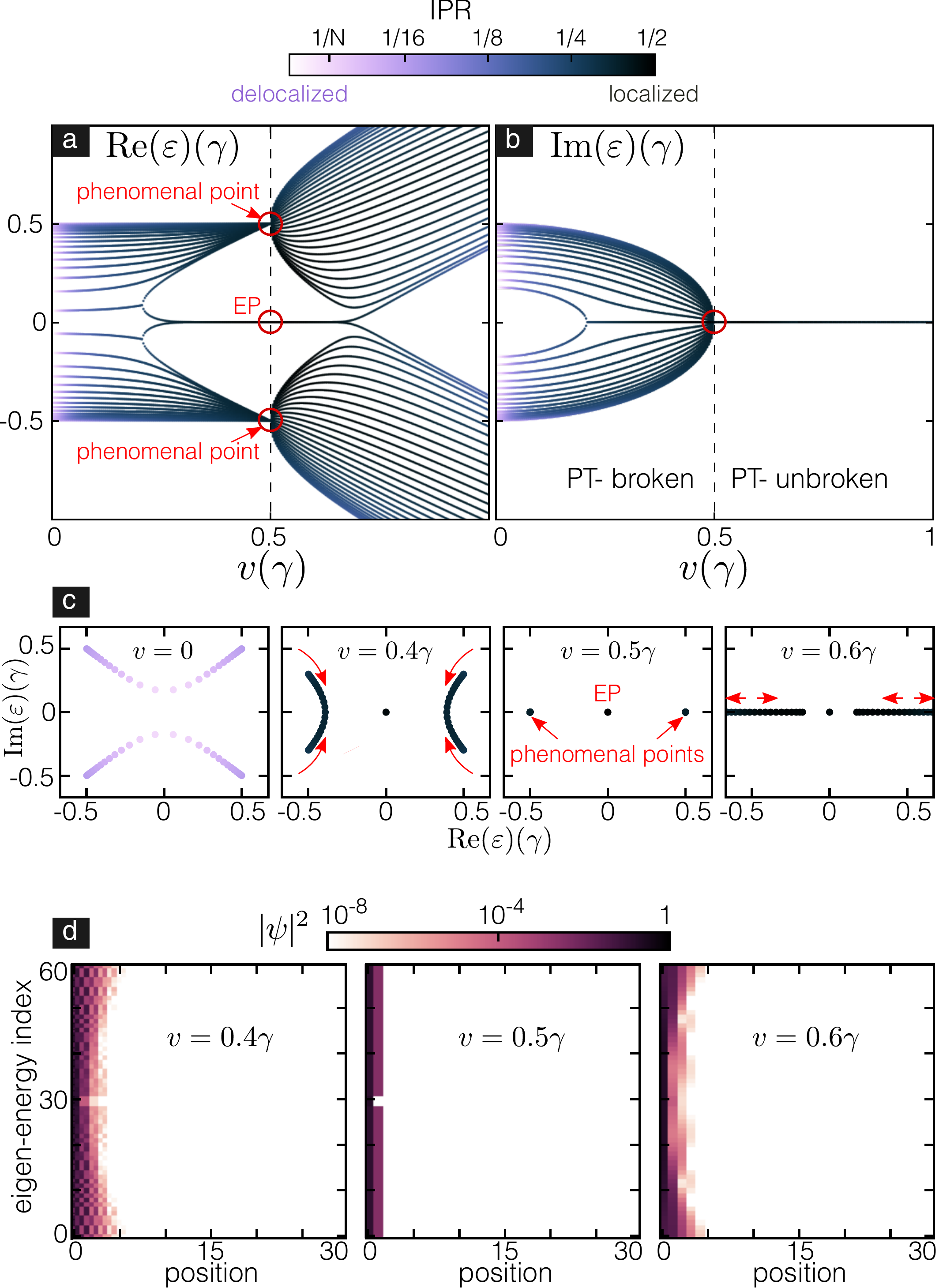}}
\subfloat{\label{Fig_2b}}
\subfloat{\label{Fig_2c}}
\subfloat{\label{Fig_2d}}
\caption{(Color online) (a) Real and (b) imaginary part of the eigen-energies obtained as a function of $v$. This corresponds to a finite system with $N=30$. The color scale shows the inverse participation ratio. (c) Detail of the evolution of the spectrum as $v$ increases from zero to some value above the singular point $v=0.5\gamma$ where exceptional (EP) and phenomenal points emerge. The red arrows indicate whether the points tend to coalesce or separate as $v$ increases. (d) Probability density associated to the eigenstates obtained for different values of $v$. Notice that all states remain localized at one edge.}
\label{Fig_2}
\end{figure}

\textit{Anomalous globally localized spectrum.--} In topological insulators, one usually devotes much less attention to the states deep in the bands than to the potentially topological edge states bridging the gap. A measure of localization of a state $\psi_{\alpha}$ is the \textit{inverse participation ratio} (IPR)~\cite{Kramer_1993,Evers2008}:
\begin{equation}
I_{\alpha}= \sum_{\textbf{r}} |\psi_{\alpha}(\textbf{r})|^4 / \left(\sum_{\textbf{r}} |\psi_{\alpha}(\textbf{r})|^2 \right)^2.
\label{ipr}
\end{equation}
The inverse of this number being roughly the average diameter of the state (in one-dimension). For extended states, $1/I_{\alpha}$ is the system's volume $L^d$. In Fig.~\ref{Fig_2}\subref{Fig_2a}-\subref{Fig_2b} the color scale encodes the IPR of each eigenstate, darker being more localized. Interestingly, close to $v=0.5\gamma$ all eigenstates, even those in the bands, are localized. This anomalous localization is maintained over a large parameter range. Indeed, the states become extended only in the immediate vicinity of $v=0$. 

One can see analytically that there are only three linearly independent eigenvectors at $v=\pm 0.5 \gamma$ where three EPs occur, one corresponding to $\varepsilon=0$ and the others to $\varepsilon=\pm r$. While the first has algebraic multiplicity $2$ and represents an EP of the usual kind, the others have an algebraic multiplicity of $N-1$ which scales with the system size, while the geometric multiplicities of all three points remains 1. \textit{Therefore, even in the large $N$ limit, a macroscopic fraction of all eigenstates coalesce onto the two states with $\varepsilon=\pm r$.} To highlight the difference with the usual EPs in regards of their large multiplicity, we call these peculiar points \textit{phenomenal points}. The change in the spectrum as $v$ changes is shown in Fig.~\ref{Fig_2}\subref{Fig_2c}.

\begin{figure}
\includegraphics[width=1.0\linewidth]{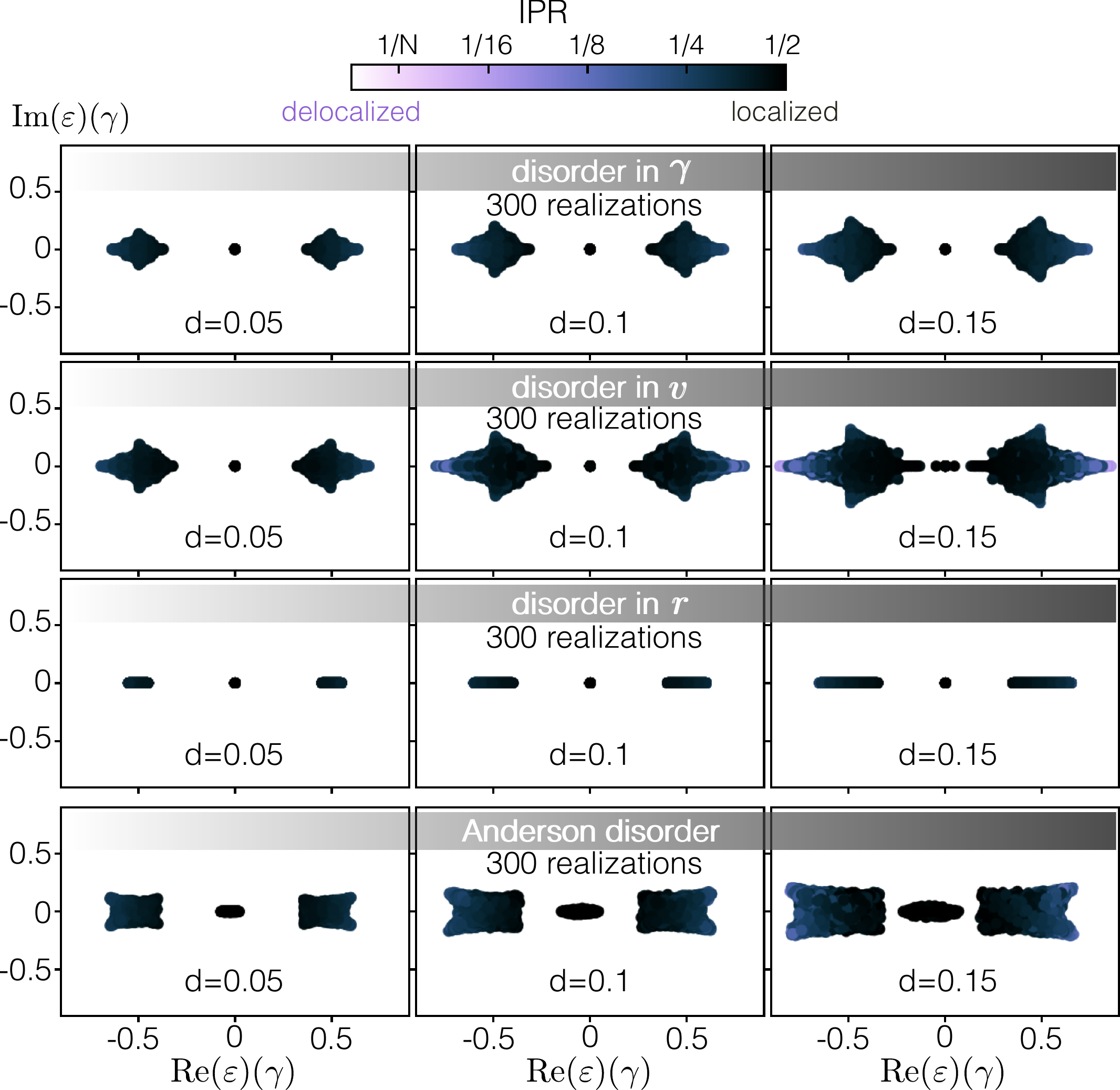}
\caption{\label{Fig_3}(Color online) Effect of increasing amounts of disorder $d$ (from left to right) in $\gamma$, $v$, $r$, and on-site Anderson disorder on the spectrum. The results for $300$ realizations of disorder and $N=30$ are superimposed. The starting point for all plots is $r=v=0.5$ and $\gamma=1$ as in the third plot in Fig.~\ref{Fig_2}c. The color scale encodes the inverse participation ratio and shows the resilience of localization against disorder.}
\end{figure}

One may wonder about the nature of the localization in such a pristine system. A closer analysis reveals that all the states are localized on the same edge as shown in Fig.~\ref{Fig_2}\subref{Fig_2d} for a few values of $v$. While isolated EPs where two states coalesce cannot have such global effect on the spectrum of a large system, the multiple coalescence near phenomenal points does. Indeed, all the three eigenstates at $v=\pm 0.5\gamma$ are localized at the same edge, thereby voiding the system of bulk states.

\textit{Robustness to disorder.--} Robustness against disorder is a characteristic feature of boundary states in a topological insulator. These states are a special subset of states with energies inside the bulk gap. In our case we will show that the \textit{global} localization of the eigenstates at one edge is resilient to moderate amounts of disorder. The Hamiltonian of Eq.~\eqref{Hamiltonian} contains three parameters: $\gamma$, $v$ and $r$ on which we introduce random disorder, $\gamma_n=\gamma+d\,\omega_n$, $v_n=v+d\, \omega_n$ and $r_n=r+d\,\omega_n$, where $n$ is the cell index, $\omega_n$ is uniformly distributed between $-1$ and $1$ and $d$ is the disorder strength. The first three rows of Fig.~\ref{Fig_3} show the results for these three types of disorder preserving the chiral symmetry. Even if the spectrum is modified, the localized nature of the states remains. Furthermore, the chiral disorder in $r$ conserves PT-symmetry unbroken as we can anticipate from Fig.~\ref{Fig_1}\subref{Fig_1b}. In Fig.~\ref{Fig_3}(bottom row), the effect of Anderson disorder breaking chiral symmetry is shown. We set the on-site energies for each site are equal to $d\omega_i$ where $i=1,\ldots,2N$ and again $\omega_n$ is uniformly distributed between -1 and 1 (Fig.~\ref{Fig_3}). Also, in this case, \textit{the global localization of the eigenstates as captured by the IPR remains quite robust to  disorder}. By robust here we mean that over a range of disorder, the localization of the spectrum is not compromised.

\textit{Massive biorthogonality around phenomenal points.--} 
We started our search looking to answer whether or not a bulk-boundary correspondence holds for this system in regards of its zero-energy edge states. Now, our interest has shifted to the full spectrum and the natural question is what causes this seemingly collective behavior that renders all eigenstates to be localized. The observed spectral properties may recall a global effect but this seems at odds with the fact that the Hamiltonian is \textit{non-interacting}. As we will see below, the key is in the non-orthogonality of the eigenstates of the non-Hermitian Hamiltonian.

\begin{figure}
\centering
\subfloat{\label{Fig_4a}\includegraphics[width=0.9\linewidth]{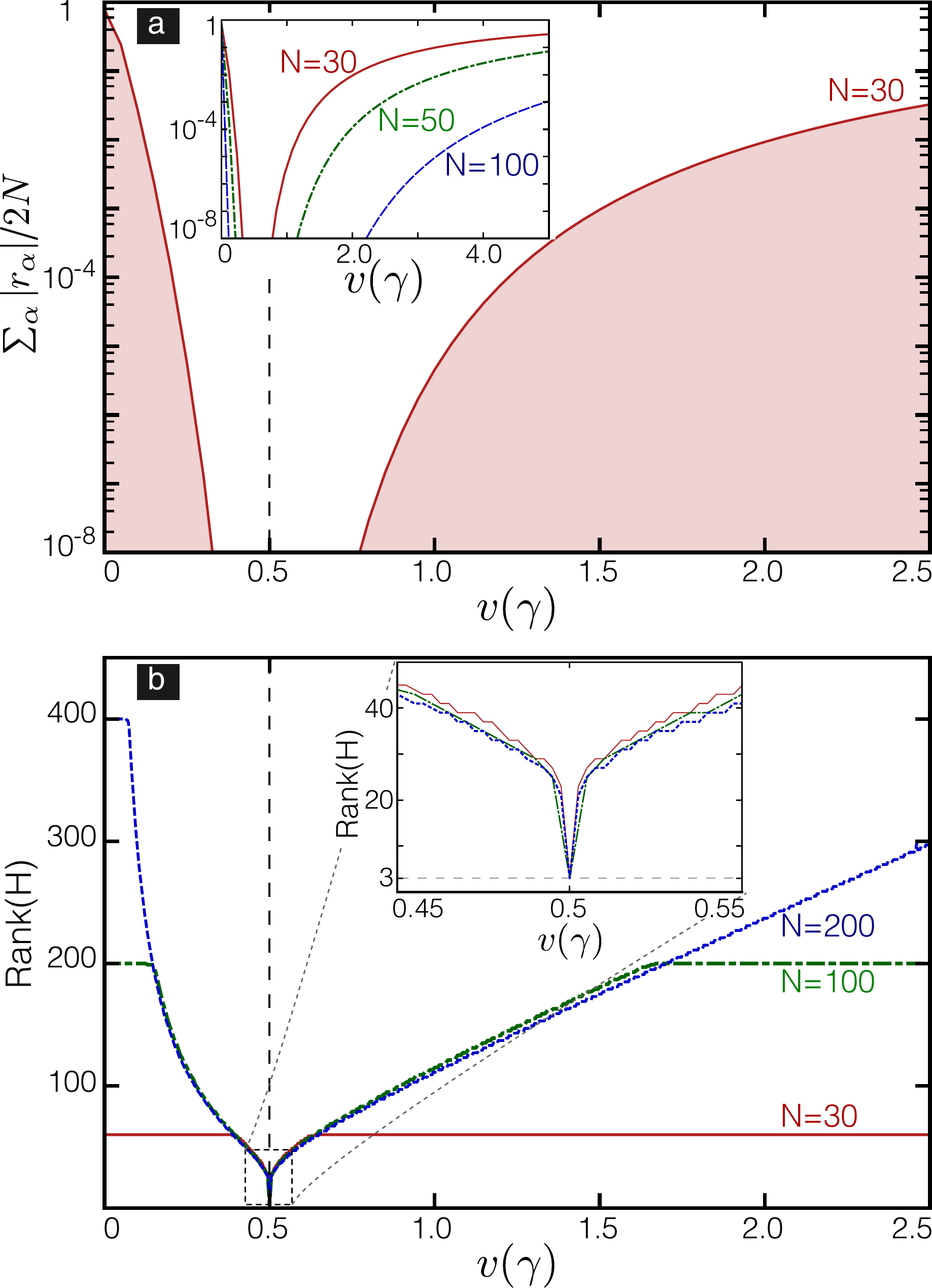}}
\subfloat{\label{Fig_4b}}
\caption{(Color online) (a) As a measure of the biorthogonality of the Hamiltonian eigenstates, we plot the sum of the absolute values of the phase rigidities $r_{\alpha}$ defined in Eq. (\ref{phase rigidity}) normalized so that it gives the value 1 when the Hamiltonian is Hermitian and 0 when all the states are biorthogonal. The main panel is for $N=30$ while the inset shows three values of $N$ over a larger range of $v$. (b) Rank of the Hamiltonian for a system containing $N$ unit cells. The inset shows a zoom over the $v$-region close to the exceptional points.}
\label{Fig_4}
\end{figure}

The right and left eigenstates and eigenenergies of ${\cal H}$ obey: ${\cal H} \ket{\psi_{\alpha}} = \varepsilon_{\alpha} \ket{\psi_{\alpha}}$ and $\bra{\phi_{\alpha}} {\cal H}= \varepsilon_{\alpha} \bra{\phi_{\alpha}}$. If ${\cal H}$ is non-Hermitian then  $\bra{\phi_{\alpha}} \neq \bra{\psi_{\alpha}}$. A measure of the eigenfunctions' biorthogonality is the phase rigidity $r_{\alpha}$ defined as~\cite{rotter_non-hermitian_2009}: 
\begin{equation}
r_{\alpha}= \frac{\braket{\phi_{\alpha} | \psi_{\alpha}}}{\braket{\psi_{\alpha} | \psi_{\alpha}}}.
\label{phase rigidity}
\end{equation}
While a Hermitian system has $r_{\alpha}=1$ for all $\alpha$, when approaching an EP $r_{\alpha}\rightarrow 0$ for the states that coalesce. Fig.~\ref{Fig_4}\subref{Fig_4a} shows $\sum_{\alpha} |r_{\alpha}| / (2N)$, with the summation extending over all states $\alpha$, thereby providing an indication of the phase rigidity of the \textit{full} spectrum. Two interesting facts arise from these data: (\textit{ii}) the biorthogonality comprehends all the eigenstates in a region around $v=0.5 \gamma$ (notice the log scale), and (\textit{ii}) the biorthogonality extends far beyond the EPs at $v=\pm 0.5 \gamma$ over a range overlapping with that found for the localized states in regards of the IPR. Strikingly, the biorthogonality becomes stronger the larger the system as the $v$ region where the phase rigidity is vanishingly small grows exponentially with system size (see inset). 

Additional insight is provided by the rank of the Hamiltonian versus $v$ as shown in Fig.~\ref{Fig_4}\subref{Fig_4b}. These results are for finite systems with $N=30$, $50$ and $100$ unit cells. The Hamiltonian has maximum rank $2N$ in the vicinity of $v=0$ and drops dramatically at $v=0.5 \gamma$ to reach the value $3$ (see inset), in agreement with our previous discussion.

As one moves away from the phenomenal points, the eigenstates change but they do it so slowly that many of them are still almost indistinguishable from each other (see Fig.~\ref{Fig_2}\subref{Fig_2d}). This is captured computing the rank of the Hamiltonian using a numerical tolerance of $10^{-8}$, see Fig.~\ref{Fig_4}\subref{Fig_4b}. The large difference between the dimension of the Hilbert space ($2N$) and this effective Hamiltonian's rank is a witness of its defectiveness.

Interestingly, the massive biorthogonality of the eigenstates also follows from the global localization around the phenomenal points:  The ket associated to a left-eigenstate ($\bra{\phi_{\alpha}}$) of ${\cal H}$, is a right-eigenstate of ${\cal H}^{\dagger}$. The spectrum of ${\cal H}^{\dagger}$ is the same as that of the original Hamiltonian but the associated eigenstates are localized on the opposite edge. Therefore, their overlap decreases exponentially with the system size leading to the rampant biorthogonality observed in Fig.~\ref{Fig_4}\subref{Fig_4a}.

We interpret our findings as an effective interaction induced by a common environment, much in the spirit of Ref.~\onlinecite{Rotter2015}. ${\cal H}$ can be decomposed into a Hermitian ${\cal H}_0=({\cal H}+{\cal H}^{\dagger})/2$ and a non-Hermitian part. The non-Hermitian part, $\Sigma=({\cal H}-{\cal H}^{\dagger})/2$ stems from a common `environment' providing for gain and loss. Since $[\Sigma,{\cal H}] \neq 0$, $\Sigma$ 
introduces the mixing among the eigenstates of $H_0$ which ultimately leads to the observed defectiveness. At $v=0.5\gamma$ the eigenvectors condense into a space of dimension $3$, independent of the dimension of the Hilbert space, $2N$. Remarkably, in spite of the decimation of the available eigenstates induced by the environment, at $v=0.5\gamma$ the $3$ remaining eigenstates become stabilized as their eigenenergies are real. This resembles the physics of resonances in scattering problems pointed out by other authors~\cite{Eleuch2016,Eleuch_PhysRevA_2017}. Thus, at the phenomenal points the ``environment'' represented by $\Sigma$ ``aligns'' the eigenstates of the Hamiltonian stabilizing them. 

\textit{Final remarks.--} Here we put forward the ordering role of exceptional points where a finite fraction of all eigenstates of a non-Hermitian Hamiltonian coalesce, the \textit{phenomenal points}. But many interesting questions remain for future study: extensions to dimensions other than one; uncovering a parent Hermitian Hamiltonian leading to the same physics; the study of phenomenal points condensing, like an Aleph~\footnote{In the short story \href{https://en.wikipedia.org/wiki/The_Aleph_(short_story)}{`The Aleph'} by Jorge Luis Borges (1945), an Aleph `is one of the points in space that contains all other points'.}, the full spectrum onto a single point. These issues may guard further surprises in a field which is already blooming~\cite{Chen2017,Hodaei2017,Rechtsman2017,Kozii2017,Xiao2017}.

\begin{acknowledgments}
LEFFT and JEBV acknowledge support from FondeCyT-Regular 1170917 (Chile) and FondeCyT-Postdoctoral 3170126.
\end{acknowledgments}

\end{document}